\documentclass[preprint,aps,showpacs]{revtex4}
\usepackage{graphicx}
\usepackage{bm}
\usepackage{amsmath}

\begin{document}

\title{Atomic Theory of Collective Excitations in Bose-Einstein Condensation
and Spontaneously Broken Gauge Symmetry}
\author{S. J. Han}

\affiliation{P.O. Box 4671, Los Alamos, NM 87544-4671\\}

\begin{abstract}

A theory of collective excitations in Bose-Einstein condensation
in a trap is developed based on the quantum Hamilton-Jacobi
equation of Bohm and the phase coherence along with the idea of
off-diagonal long range order of Penrose and Onsager. First, we
show that a free surface behaves like a normal fluid - a breakdown
of superfluidity. Second, inside the free surface it is shown that
the spectrum of phonons is of the form $\omega=ck$ scaled with the
external potential, where the speed of (first) sound, $c=[4\pi
a\rho\hbar^{2}]^{1/2}/M$ and $k$ is the wave number. Third, in the
limit $a\rightarrow 0$, the hard spheres in the Bose-Einstein
condensation collapse to a close-packed classical lattice with the
zero-point vibrational motion about fixed points.

\end{abstract}

\pacs{03.75.Fi, 03.65.-w, 03.65Sq, 05.30-d}

\maketitle
%\{widetext}

The new Bose-Einstein condensation (BEC) in a trap is one of the
remarkable phenomena exhibited by a macroscopic system of Bose gas
at low temperature \cite{Lieb02}. In his seminal paper on a
theoretical model of a Bose gas \cite{Bog47}, Bogoliubov has laid
out the basis for much of our understanding of superfluidity of a
system of weakly interacting Bose particles. The concept of BEC is
generalized to strongly interacting systems by Penrose and Onsager
based on the notion of off-diagonal long range order (ODLRO)
\cite{Penrose56} which gives rise to superfluidity in He II. Many
properties of an interacting Bose system can be understood in
terms of independent quasi-particle excitations (phonons and
rotons) of the system. These excitations are essential for a
complete understanding of certain properties of the interacting
Bose system, in particular, the longitudinal excitation spectrum
in the phonon regime is crucial for a quantitative experimental
confirmation of BEC in a trap.

However, the Bogoliubov theory of collective excitations (first
sound) differs so basically in its formulation from that of
Feynman's atomic theory of two-fluid model \cite{Feynman53} and
Landau's two-fluid model \cite{Landau41} that the problem of the
collective excitations in BEC represents the major stumbling block
in our attempt to confirm the new BEC in terms of the Bogoliubov
theory. For instance, the new Bose condensates in a trap are
extremely small in size ({\it i.e.,} a liquid droplet of tens of
microns in radius), and unlike He II they are highly compressible
and have a nonuniform density profile. In addition, the new BEC is
dilute (mean particle density $\bar\rho$ satisfies $\bar\rho
a^{3}\ll 1$) \cite{Fetter01}. It is therefore obvious that the
formal extension of Bogoliubov's theory to the new BEC is
difficult on account of its inhomogeneous density profile.
Furthermore, it is impossible in our study of the collective
excitations to make use of the elegant quantum field theory
technique developed in many-body theory which is valid only for a
uniform system \cite{Bog47,note1}.

The new BEC is, however, the simplest of all condensed many-body
entities from the microscopic point of view. The atoms in a trap
may be treated as a weakly interacting Bose gas which can be
accurately described by a self-consistent Hartree equation with
the well-known scattering length in the hard sphere approximation
for a pair-interaction \cite{Lee57}.

On the experimental side, the first systematic experimental study
on the collective excitation was recently carried out in a long
$^{87}$Rb Bose condensate in which the sound wave propagation can
be treated approximately as a $1$-D problem \cite{Steinhauer02}.
Yet the experimental data yield a qualitatively correct dispersion
curve of the Bogoliubov spectrum $\omega=ck$ with the estimated
speed of first sound $c_{eff}=2.0\pm 0.1 mm sec^{-1}$. Before the
ideas of the new BEC \cite{Lieb02} in a trap become tenable, it
will be necessary to show experimentally that the collective
excitation spectrum in the new BEC indeed agrees well with the
Bogoliubov spectrum with the proper geometrical corrections
\cite{Bog47,Penrose56}.

Theory, in the absence of credible experiments, could wander off
down a blind alley \cite{Fetter01}. And yet it is the theoretical
insight that could unravel how Nature creates such a unique object
with a peculiar property. Our analysis shows a remarkable new
phenomenon - a breakdown of superfluidity at the free surface of
BEC in a trap. This may be interpreted by broken gauge symmetry
\cite{Anderson66,BenLee73, Weinberg96} and is precisely the reason
why the speed of first sound was difficult to measure
\cite{Steinhauer02}.

The object of this paper is, first, to show how Bohm's quantum
theory \cite{Bohm52} can be applied to the problem of collective
excitations in BEC in such a way as to incorporate Feynman's
atomic theory of two-fluid model \cite{Feynman53} and the concept
of phase coherence in ODLRO, and, secondly, to present a
demonstration that symmetry breaking is taking place at the free
surface with the first-order symmetry breaking perturbations in
Lagrangian coordinates \cite{BFKK58,Weinberg67,SJHan82}. It is
shown here for the first time that the dispersion relation for the
surface wave is a manifestation of the broken symmetry at the free
surface. The symmetry breaking also accompanies massless phonons
(Nambu-Goldstone bosons) \cite{Weinberg96}, the spectrum of which
is shown to be the same in form as that of the Bogoliubov
dispersion relation $\omega=ck$ scaled by the external trapping
force with a small geometrical correction, where the speed of
(first) sound, $c=[4\pi a \rho \hbar^{2}]^{1/2}/M$ and $k$ is the
wave number \cite{Bog47}.

Since the perturbation techniques of quantum field theory
\cite{Bog47,note1} cannot be applied to a finite, spatially
inhomogeneous system, we must find a new perturbation method that
yields the Bogoliubov spectrum of collective excitations in the
new BEC in a trap. Here we present straightforward, but
mathematically precise, first-order perturbations by particle
displacements in Lagrangian coordinates to an ensemble of particle
trajectories defined by the condition of BEC that automatically
include all the excited modes which tend to be otherwise omitted.
In the present analysis, the problem of collective excitations
reduces to a boundary-value problem for a BEC droplet in a trap
bounded by a free surface. The present analysis also gives a good
account of the possible formation of a close-packed lattice of a
trapped Bose gas.

The essential point of ODLRO is that BEC (and superfluidity) be
described as a state in which the density matrix can be factorized
as
\begin{equation}
\rho(\bm{x},\bm{x}^{\prime})=\psi^{\dagger}(\bm{x})\psi(\bm{x}^{\prime})+
\gamma(|\bm{x}-\bm{x}^{\prime}| \label{Odlro},
\end{equation}
where $\gamma\rightarrow 0$ as $|\bm{x}-\bm{x}^{\prime}|
\rightarrow \infty$. Here the single particle wave function $
\psi(\bm{x})$ (mean field) represents the Bose condensed state
\cite{Anderson58} and $\int\psi^{\dagger}\psi d\bm{x}=N$, where $N$
is the number of particles in BEC. With the the hard sphere
approximation of the inter-particle repulsive force for Bose
particles \cite{Lee57}, one can show that the mean field satisfies
the nonlinear Schr\"{o}dinger equation (Gross-Pitaevskii),
\begin{equation}
i\hbar\frac{\partial\psi}{\partial t}=-\\
\frac{\hbar^{2}}{2M}\nabla^{2}\psi + V({\bm x})_{ext} +g|\psi|^{2}\psi,\\
\label{Cat2}
\end{equation}
where $g=4\pi\hbar^{2}a/M$ and $a$ is the s-wave scattering
length. Eq.~(\ref{Cat2}) is a self-consistent Hartree equation for
the Bose condensed wave function. This short-range interaction
brings about the Bose condensed state also ensures that the system
possess the longitudinal collective excitations (phonons) in BEC.
It should be noted that the nonlinear term $|\psi|^{2}$ is
invariant under a $U(1)$ group transformation. Hence we can apply
Bohm's interpretation of quantum theory to Eq.~(\ref{Cat2}) with
the understanding that it is a part of the potential defined in
his quantum theory.

We write the equations for the ensemble average energy in the
usual quantum theory \cite{Bohm52,Aharonov63,Lieb98}:
\begin{subequations}
\label{allequations} \label{ground}
\begin{eqnarray}
{\cal H} =\int \psi^{\dagger}\left(-\frac{\hbar^{2}}{2M}\nabla^{2}
+
V(\bm{x})_{ext} +\frac{g}{2}|\psi|^{2}\right)\psi d\bm{x}, \label{grounda}\\
{\cal E}_{ave} =\int\left(\frac{\hbar^{2}}{2M}|\nabla\psi|^{2} +
V(\bm {x})_{ext}|\psi|^{2}+\frac{g}{2}|\psi|^{4}\right) d\bm{x},
\int\psi^{\dagger}\psi d\bm{x}=N.\label{groundb}
\end{eqnarray}
\end{subequations}

At this point it is possible to show that the ground state density
is given in terms of an external potential and the chemical
potential upon minimizing the energy functional Eq.~(\ref{groundb})
for a fixed number of particles in a trap with the condition
$\bm{p}=0$ (Penrose-Onsager criterion for BEC):
\begin{equation}
\rho(\bm{x})=|\psi(\bm{x})|^{2}=\frac{M}{4\pi\hbar^{2}a}[\mu
-V(\bm{x})_{ext}], \label{Ground}
\end{equation}
where $\mu$ is a Lagrangian multiplier and is the chemical
potential. It should be noted that the ground-state wave function
Eq.~(\ref{Ground}) has a nodal surface at the boundary where the
symmetry of the Bose system breaks down.

To briefly recapitulate Bohm's interpretation of quantum theory
\cite{Bohm52}, we write the wave function in the form
$\psi(r,t)=f(\bm{r},t)exp[\frac{i}{\hbar}S(r,t)]=
\rho^{1/2}(\bm{r},t)exp[\frac{i}{\hbar}S(r,t)]$, where $S(r,t)$ is
an action (or a phase)\cite{Aharonov63}. We then rewrite
Eq.~(\ref{Cat2}) to obtain,
\begin{subequations}
\label{allequations} \label{Mean}
\begin{equation}
\frac{\partial\rho}{\partial t}+\bm{\nabla}\cdot(\rho\frac{\bm\nabla S}{M})=0\\
\label{Density}
\end{equation}
\begin{eqnarray}
 \frac{\partial S}{\partial t} + \frac{(\bm\nabla S)^{2}}{2M} +
V(\bm{x}) - \frac{\hbar^{2}}{4M}[\frac{\nabla^{2}\rho}{\rho}-
\frac{1}{2}\frac{(\nabla\rho)^{2}}{\rho^{2}}]=0, \label{QHJ}
\end{eqnarray}
\end{subequations}
where $ V(\bm{x})=V(\bm{x})_{ext}+g\rho$. The last term $U_{eqmp}$
of Eq.~(\ref{QHJ}) is the effective quantum-mechanical potential
(EQMP) which plays an important role in an inhomogeneous
condensate and fluctuates near the free surface as will be shown
shortly.

From the point of view of macroscopic physics, the coordinates and
momenta of an individual atoms are hidden variables, which in a
macroscopic system manifest themselves only as statistical
averages. In the following we show this interpretation by deriving
the dynamical equations from Eqs.~(\ref{Mean}) with
$\bm{v}(\bm{r}_{0},t)=\bm{\nabla}S(\bm{r}_{0},t)/M$ for a single
particle in condensation represented by the mean field in
Eq.~(\ref{Cat2}).

In a dilute Bose gas in a trap, the force derived from the
potential $V(\bm{x})_{ext}$ balances the repulsive force due to $g
\rho(\bm{x})$ to create BEC for which ${\bm p}(\bm{x})=0$. Hence
in the limit $\hbar\rightarrow 0$ or with the less stringent
condition ${\bm\nabla}\rho=0$ in a homogeneous medium for which
$U_{eqmp}=0$, $S(\bm{x},t)$ is a solution of the Hamilton-Jacobi
equation in Bohm's theory \cite{note2}. Eq.~(\ref{Density})
expresses a probability density of a statistical ensemble for a
system. Eq.~(\ref{QHJ}) implies, however, that the particle moves
under the action of the force which is not entirely derivable from
the potential $V(\bm{x})$, but which also obtains contributions
from $U(\bm{x})_{eqmp}$. It is the essence of Bohm's
interpretation of quantum theory. In the classical description,
the particle follows a classical orbit determined by the
Hamilton-Jacobi equation. Eqs.~(\ref{Mean}) is a function of
macroscopic dynamical variables and suggest us how we may choose
the most stable particle trajectories ({\it classical orbits})
about which we may linearize the equations of motion.

The first step in the development of a theory of collective
excitations in BEC is to associate with a single particle wave
function in ODLRO with precisely definable, symmetry breaking
perturbations to the first order. Since external perturbations are
always in the classical domain, the problem of collective
excitations is reduced to finding atomic displacements that are
consistent with the atomic theory of the two-fluid model of He II
by Feynman in which the exact partition function is given as an
integral over particle trajectories, using his space-time approach
to quantum mechanics \cite{Feynman53}. We wish to stress that the
collective excitation spectrum is also a test of the perturbation
method for calculating the longitudinal excitations of (first)
sound in BEC in a trap.

Our perturbation method is based on the following physical
picture: the collective coordinates are attached to the particles
in motion as the phase coherent sound wave propagates from the
center to the surface of BEC, and that the sound wave must also
satisfy the prescribed initial-boundary conditions at the surface.
The surface excitation involves the breaking up of phase coherence
over the entire surface and is accompanied by a dissipation
process of the sound wave. This gives rise a surface energy to
conserve the total energy of an isolated system.

To describe this, we introduce symmetry breaking perturbations to
the particles in BEC in Lagrangian coordinates
\cite{BFKK58,Weinberg67,SJHan82}, which can be defined as
$\bm{x}=\bm{x}_{0}+\bm{\xi}(\bm{x}_{0},t)$. Here
$\bm{\xi}(\bm{x}_{0},t)$ is a function of the unperturbed position
of a particle in the condensation, and remains attached to the
particle in the condensate as it moves. This semiclassical
perturbation method is in fact a well-defined accurate quantum
mechanical approximation scheme which yields the Bogoliubov
spectrum for the collective excitations \cite{Bog47,SJHan2}.

At this stage we find it more convenient to work with the familiar
dynamical equations instead of Eqs.~(\ref{Mean}). We also note
that the use of the Hamilton-Jacobi equation in solving for the
motion of a particle is only matter of convenience.

If we write $\bm {v}(\bm{x}_{0},t)=\bm{\nabla}S(\bm{x}_{0},t)/M$ -
(a solution of the Hamilton-Jacobi equation), we then obtain the
following three dynamical equations from Eqs.~(\ref{Ground}) and
(\ref{Mean}):

the equation of motion for a single particle,
\begin{equation}
M(\frac{\partial}{\partial t}\bm{v}+\bm{v}\cdot \bm{\nabla} \bm
{v})=-\bm{\nabla}\mu, \label{Motion0}
\end{equation}
the equation of continuity,
\begin{equation}
\frac{\partial}{\partial t}\rho+\bm{\nabla}\cdot(\rho\bm{v})=0,%
\label{Cont0}
\end{equation}
where $\rho$ is henceforth interpreted as the number density,

and the equation of state,
\begin{equation}
\mu(\bm{r}, t)=\mu_{\text{loc}}[\rho(\bm{r}, t)]+V_{\text{ext}},
\label{Eqs0}
\end{equation}
where $\mu_{\text{loc}}[\rho(\bf r, t)]$ is the chemical potential
in the local density approximation \cite{Oliva89}.

Next we expand $S$, $\bm{v}$, and $\rho$ to the first-order,
\begin{subequations}
\label{allequations} \label{First}
\begin{eqnarray}
S(\bm{r},t)=S(\bm{r}_{0},t) +
\bm{\xi}\cdot\bm{\nabla}_{0}S(\bm{r}_{0},t)\\
\label{Firsta} \bm{v}(\bm {x},t)=\frac {\partial}{\partial
t}\bm{\xi}
\label{Firstb}\\
\rho(\bm{x},t)=\rho(\bm{x}_{0})-
\bm{\nabla}_{0}\cdot[\rho(\bm{x}_{0})\bm{\xi}]=
\rho(\bm{x}_{0})+\delta \rho \label{Firstc},
\end{eqnarray}
\end{subequations}
where $\bm{p}(\bm{x}_{0})=0$ in BEC and $\bm{\nabla}_{0}$ denotes
the partial derivative with respect to $\bm{x}_{0}$ with
$\bm{\nabla}\rightarrow
\bm{\nabla}_{0}-\bm{\nabla}_{0}\bm{\xi}\cdot\bm{\nabla}_{0}$.
Eq.~(\ref{Firstc}) was derived by substituting Eq.~(\ref{Firstb})
to Eq.~(\ref{Cont0}) with $\bm{\nabla}S(\bm{x},t)=
\bm{p(\bm{x},t)}$ without taking into account of EQMP in
Eq.~(\ref{QHJ}), then integrating over time.

Now it is easy to express from Eqs.~(\ref{First}) the boundary
conditions in terms of ${\bf\xi}$ by the usual definition of
incompressibility and irrotational motion of a fluid at the free
surface in BEC \cite{Landau59,Lamb45}; they are given
mathematically by ${\bf\nabla}\cdot{\bf\xi}=0$ and
${\bf\nabla}\times{\bf\xi}=0$.

The phase coherence is essential to relate the Bose condensed wave
function (mean field) in ODLRO to a many-body ground-state wave
function, $\bm{\xi}\cdot\bm{\nabla}S(\bm{x}_{0},t)
=\sum_{i}\bm{\xi}_{i}\cdot\bm{\bm{\nabla}_{i}}S_{0,i}(\bm{x}_{0,i},t)$
in our analysis, where $S_{0,i}(\bm{x}_{0,i},t)$ is the phase of a
single Bose particle in the system, and is a necessary and
sufficient condition for phase coherence. Hence the collective
excitations we study in this paper are phase coherent waves.

The basis of this paper is the set of linear equations,
Eqs.~(\ref{Motion0}), (\ref{Cont0}), (\ref{Eqs0}), and
(\ref{First}) that describe the collective excitations in BEC. It
should be remarked that $\psi$ can support the collective
excitations (phonons) only if it obeys the nonlinear
Schr\"{o}dinger equation for which we take
$\bm{\nabla}S(\bm{x}_{0},t)= \bm{p(\bm{x}_{0},t)}$ as a solution
of the Hamilton-Jacobi  just as the nonlinear Maxwell equations
describe radiation processes (photons).

By virtue of Eq.~(\ref{Firstc}), the chemical potential may be
expanded as \cite{Oliva89}
\begin{equation}
\mu(\bm{x}, t) = \mu_{\text{loc}}(\bm{x}_{0}) + V_{\text{
ext}}-\frac{\partial}{\partial \rho}(\mu_{\text
{loc}})[\bm{\nabla}_{0}\cdot(\rho_{0}\bm{\xi})]. \label{Eqs1}
\end{equation}

In BEC, $\bm{p}(\bm{x}_{0})=0$, and the equation of motion gives
\begin{equation}
\mu_{\text{loc0}}(\rho_{0}(\bm{r}_{0}))+V_{\text{ext}}=constant
=\mu_{\text{loc0}}[\rho_{0}(0)]=\mu_{0}. %
\label{Eqs2}
\end{equation}

If we now set $V_{\text{ext}}=\omega_{0}^{2}r^{2}/2$, the density
profile describes a spherically symmetric, nonuniform
Bose-Einstein condensation (SSNBEC). This is the problem from
which our investigation started. For the special example of
collective excitations in BEC in trap, we limit our analysis to
the problem of SSNBEC.

In order to describe the collective excitations, we linearize
Eqs.~(\ref{Motion0}) along with Eq.~(\ref{Firstb}),
Eq.~(\ref{Firstc}), Eq.~(\ref{Eqs1}), and Eq.~(\ref{Eqs2}). It is
straightforward algebra, though somewhat tedious, to arrive at the
first-order equation of motion in $\bm{\xi}$. The result is

\begin{equation}
\frac{\partial^{2}}{\partial t^{2}}\bm
{\xi}=\frac{\mu_{0}}{M}\bm{\nabla}\sigma-\omega_{0}^{2}[(\bm{
\xi}\cdot\bm{\nabla})\bm{r}+(\bm {r}
\cdot\bm{\nabla}\bm{\xi})]\\
-\omega_{0}^{2}[\sigma\bm {r}
+\frac{1}{2}r^{2}\bm{\nabla}\sigma].%
\label{Main}
\end{equation}
Here we have taken $V_{ext}(r)=M\omega_{0}^{2}r^{2}/2$, $\partial
\mu_{loc}/\partial \rho =4\pi\hbar^{2}a/M$,  $\sigma={\bf
\nabla}\cdot{\bf\xi}$, and have also dropped the subscript in
$r_{0}$.

Eq.~(\ref{Main}) is a typical second-order (inhomogeneous) partial
differential equation subject to the initial-boundary conditions
and has two particular solutions, one corresponding to the surface
waves and the other to longitudinal sound waves. The general
solution to Eq.~(\ref{Main}) is a combination of the two
solutions.

We begin with the surface phenomena, and shall study later the
more complex problem of the longitudinal sound waves in SSNBEC. We
limit the surface waves by imposing the boundary conditions
${\bf\nabla}\cdot{\bf\xi}=0$ and ${\bf\nabla}\times{\bf\xi}=0$,
which simplifies the algebra considerably. This implies that we
may solve for the potential flow as $\nabla^{2}\chi=0$ and $\bm
{\xi}_{\text{s}}=-\bm {\nabla}\chi$. Here the subscript $s$ stands
for the surface waves. The solution for $\chi$ is given by
\begin{equation}
\chi(r, t)=
\sum_{\ell,m}[\chi_{+}^{\ell}(t)r^{\ell}+\chi_{-}^{\ell}(t)r^{-(\ell+1)}]
 Y_{\ell,m}(\theta, \phi),
\end{equation}
where we may set $\chi_{-}^{\ell}(t)=0$ for a quantum liquid
droplet. We may now expand $\bm{\xi}_{s}$ in terms of three
orthogonal vector spherical harmonics \cite{SJHan1},
\begin {equation}
\bm{\xi}_{s}=\sum_{\ell,m}[\xi_{1s}^{\ell,m}(r,t)\bm{
a}_{1}+\xi_{2s}^{\ell,m}(r,t)\bm{a}_{2}+\xi_{3s}^{\ell,m}(r,t)\bm{
a}_{3}]. \label{Surf1}
\end {equation}
Here we have defined the three vector spherical harmonics as

\[\bm {a}_{1}=\bm {e}_{r}Y_{\ell,m}(\theta,\phi),~ \bm{a}_{2}=r \bm
{\nabla}Y_{\ell,m}(\theta,\phi), ~\bm {a}_{3}=\bm {r}\times \bm
{\nabla}Y_{\ell,m}(\theta,\phi).\]

It is now only a matter of elementary algebra, by substituting
Eq.~(\ref{Surf1}) into Eq.~(\ref{Main}) to obtain a dispersion
relation,
\begin {equation}
\omega^{2}_{surf}=\ell\,\omega_{0}^{2}. \label{Surf2}
\end{equation}

Here we have taken $\chi_{+}(t)=e^{i\omega t}$. It is at once
apparent that the dispersion relation for the surface waves is
independent of the internal dynamics of the trapped Bose gas ({\it
e.g.,} the pair interaction potential). Another way of looking at
the property of the free surface is to note that the surface waves
are driven by the action of external perturbations over the
surface of a spherical droplet in equilibrium, similar to that of
gravity waves on the surface of a fluid in equilibrium in a
gravitational field \cite{Landau59,Lamb45}. More importantly,
\textit {the dispersion relation is independent of $\hbar$ which
implies that the free surface behaves like a classical fluid}.

It is, therefore, evident that Eq.~(\ref{Surf2}) is the first
definite proof that the free surface of a superfluid behaves like
a normal fluid; it is a peculiarly universal property of the free
surface of a superfluid under an external field. Furthermore,
Eq.~(\ref{Surf2}) is also an excellent example of broken symmetry
in Nature \cite{Anderson66}. The breakdown of superfluidity has
been already observed at the vortex core in He II
\cite{Reif64,Glaberson68}. The breakdown of superfluidity at the
free surface in rotating He II \cite{SJHan2} also resolves a long
standing puzzle in Landau's two fluid model in connection with the
question of why a superfluid component rotates with the normal
fluid at the free surface \cite{Landau41,Osborne50, Meservey64}.

Since in practice the free surface is never confined to a strictly
mathematical surface, the thin surface layer which also contains
phonons and particles driven by EQMP,

\begin{equation}
U_{eqmp}=-\frac{\hbar^{2}}{4M}[\frac{\nabla^{2}\rho}{\rho}-
\frac{1}{2}\frac{(\nabla\rho)^{2}}{\rho^{2}}]
=-\frac{\hbar^{2}}{M}\frac{\nabla^{2}f}{f}. \label{Eqmp}
\end{equation}

Thus the particles experience a force from $U_{eqmp}$ which
fluctuates with the momentum $\bm{p}=\bm{\nabla}S$ and energy of a
particle near the surface with the degree of divergence
$M\omega_{0}^{2}\mu_{0}/D^{2}$ as
$D=[\mu_{0}-(1/2)M\omega^{2}_{0}r^{2}]\rightarrow 0$ at the free
surface as emphasized by Bohm \cite{Bohm52}. $U_{eqmp}$,
therefore, breaks up phase coherence in the collective excitations
in the surface layer.

In our model of an imperfect Bose gas, between each pair of
particles there is a hard sphere repulsion of range $a$ and no
long-range interaction. This pair-interaction ensures that the
system possesses the longitudinal collective excitations (phonons)
in BEC. Thus the problem of collective excitations in SSNBEC
should be considered as a spherical longitudinal, phase-coherent
sound wave propagation. In the phonon regime, the dispersion
relation for phonons must be of the form $\omega=ck$ scaled with
the external trapping force, where the speed of first sound,
$c=[4\pi a\rho\hbar^{2}]^{1/2}/M$, and $k$ is the wave number.
Because of our approximation for the pair-interactions of Bose
particles in the system, the presence of the s-wave scattering
length $a$ in the speed of (first) sound wave $c$ is essential.
Also this approximation should be valid so long as the condition
$\bar\rho a^{3}\ll 1$ is met, a condition which restricts the
discussion to the region of temperature near absolute zero, which
explains why there cannot be roton excitations in BEC in a trap.

Now we are ready to define the compressibility of the fluid as
$\sigma=\bm{\nabla}\cdot{\bm {\xi}}$ from Eq.~(\ref{Firstc}), and
then its spectrum can be obtained from Eq.~(\ref{Main}) by taking
the divergence on both sides. After straightforward algebra with
the help of the vector identity ${\bf\nabla}\cdot[({\bf
r}\cdot{\bf\nabla}){\bf\xi}]={\bf r}\cdot{\bf
\nabla}\sigma+\sigma$ and the condition of superfluidity,
${\bf\nabla}\times{\bf\xi}=0$, we obtain
\begin{equation}
\frac{\partial^{2}}{\partial t^{2}}
\sigma(r,t)=\frac{1}{2}\omega_{0}^{2}(\alpha^{2}-r^{2})\nabla^{2}\sigma-
3\omega_{0}^{2}r\frac{\partial}{\partial
r}\sigma-5\omega_{0}^{2}\sigma,
\end{equation}
where $\alpha^{2}=8\pi\hbar^{2}a \rho_{0}(0)/(M\omega_{0})^{2}$.

The solution of the equation is not entirely trivial. Writing
$\sigma(\bf r, t)$= $S(t)W(r)Y_{\ell,m}(\theta, \phi)$, we obtain
the variable separated equations:
\begin{equation}
\frac{d^{2}}{dt^{2}}S(t)+\lambda_{n} S(t)=0%
\label{Time}
\end{equation}
and
\begin{eqnarray}
(\alpha^{2}-r^{2})[\frac{1}{r}\frac{d^{2}}{dr^{2}}(rW_{n}(r))-
\frac{\ell(\ell+1)}{r^{2}}W_{n}(r)]\nonumber\\
-6r\frac{d}{dr}W_{n}(r)\nonumber\\
+(2\lambda_{n}/\omega_{0}^{2}-10)W_{n}(r)=0,%
\label{Space}
\end{eqnarray}
where $\lambda_{n}$ is a constant of separation. The eigenvalues
are determined by Eq.~(\ref{Space}), and by the boundary
conditions on $\sigma(r,t)$, {\it i.e.,} $\sigma(r,t)=0$ at the
free surface and the origin. Here we tacitly assume an infinitely
small point source at the origin to drive an outgoing spherical
sound wave \cite{Landau59}. The speed of first sound $c=[4\pi a
\rho(r)\hbar^{2}]^{1/2}/M $ is the maximum at the origin and
becomes zero at the free surface. To see this is indeed the case
we recall that the surface layer is no longer a superfluid but is
a normal fluid. Hence the phonons can interact with the normal
fluid and, upon the dissipation process, give rise to a surface
energy.

The solution of Eq.~(\ref{Time}) then gives the dispersion
relation for the collective excitations with the eigenvalues that
are a function of the s-wave scattering length, the radial trap
frequency, the peak density at the center of the trap, {\it i.e.,}
the speed of first sound $c_{ctr}=[4\pi a
\rho(0)\hbar^{2}]^{1/2}/M$, the trapping frequency, and the wave
number with $k_{\theta}\simeq \ell/r$. To show this explicitly, we
transform Eq.~(\ref{Space}) to the Sturm-Liouville problem and
obtain the eigenvalues in terms of a complete set of functions
with the orthogonal properties:
\begin{equation}
 \acute{\lambda_{n}}=\int_{0}^{b}r^{2}(\alpha^{2}-r^{2})^{3}
[(\frac{d}{dr}W_{n})^{2}+
\frac{\ell(\ell+1)}{r^{2}}W_{n}^{2}(r)]dr %
\label{Eigen}
\end{equation}
 and with
\begin{equation}
\int_{0}^{b}r^{2}(\alpha^{2}-r^{2})^{2}W_{m}(r)W_{n}(r)dr
=\delta_{m,n}.
\label{Norm}
\end{equation}
Here $\acute{\lambda_{n}}=(2\lambda_{n}/\omega_{0}^{2}-10)$ and
$b=\alpha=[8\pi a\hbar^{2}\rho(0)]^{1/2}/(M\omega_{0})$ is the
radius of the condensate. This procedure, though somewhat a
digression, is the only way we can explicitly show the functional
relation between the eigenvalues and the speed of first sound.

Equations ~(\ref{Eigen})-(\ref{Norm}) exhibit the main feature of
our method: that the eigenvalues are, indeed, a function of the
speed of (first) sound, $b=[\sqrt{2}/\omega_{0}]\,c_{ctr}$, where
the sound speed $c_{ctr}=[4\pi a \rho(0)\hbar^{2}]^{1/2}/M$, and
the trapping frequency $\omega_{0}$, for all values of the angular
momentum $\ell$. This method of presentation also shows why the
excitation spectrum has not shown the dependence, or lack thereof,
the speed of (first) sound in previous studies \cite{Fetter01}.
The functional relation of the speed of (first) sound and the wave
number in the excitation spectrum is also clear in this integral
representation. More importantly, it is apparent that the
perturbation method by Lagrangian displacement vectors
\cite{BFKK58,Weinberg67,SJHan82} is in fact a well-defined quantum
mechanical approximation scheme.

Another important point is that the eigenvalue is a function of
the speed of first sound not at the free surface but at the center
of SSNBEC. This has a simple physical interpretation: an outgoing
spherical sound wave can travel with little energy loss toward the
free surface where the sound wave (phonons) interacts only with
the normal fluid of the surface layer and completely dissipates
giving rise to a surface energy. It is precisely the nature of a
superfluid that it cannot interact with phonons, but the phonons
will interact with a normal fluid as Mott emphasized
\cite{Mott49}. Therefore, the phenomenon of the fluctuation of
particles due to $U_{eqmp}$ and the dissipation of sound waves at
the surface layer can be understood in terms of Kubo's
fluctuation-dissipation theory \cite{Kubo57} and satisfies the
conservation of energy for an isolated system. We thus see that
the particles in the surface layer obey the Boltzman statistics
just like rotons because of the elevated particle energy. Hence
the surface layer is a normal fluid in every respect
\cite{Landau41}.

The analytical expression Eq.~(\ref{Eigen}), in principle, gives a
complete solution to the eigenvalue problem. A moment's thought
shows, however, that the expression for $W_{n}$ is too complicated
to evaluate the eigenvalues by this approach. Here we follow a
different, but equivalent, approach to the eigenvalue problem;
that is, we shall solve Eq.~(\ref{Space}) as an eigenvalue
equation. With the substitution
$W_{n}(r)=r^{\pm(\ell+1/2)-1/2}Z_{n}(r)$ together with
$x=r^{2}/\alpha^{2}$, we obtain
\begin{equation}
x(1-x)\frac{d^{2}}{dx^{2}}Z_{n}+[c-(a+b+1)x] \frac{d}{dx}Z_{n}
-abZ_{n}=0
\label{Last}
\end{equation}
This is just the Gauss differential equation \cite{WW63} with
$c=\pm(\ell+1/2)+1$, $a+b=\pm(\ell+1/2)+3$, and
$ab=(1/4)\,\acute{\lambda}_{n}-6\,[\pm(\ell+1/2)-1/2]$.

Eq.~(\ref{Last}) has regular singular points at $x=0$, $x=1$ and
$x=\infty$. Its solution is the hypergeometric function, which is
analytic in the complex plane with a cut from $1$ to $\infty$ along
the real axis \cite{WW63}.

Since we are interested in the low-lying excited states (phonons),
it is only necessary to find the smallest eigenvalues in the
differential equation. This is consistent with Feynman's picture
of phonons - a sound (longitudinal) wave without nodes
\cite{Feynman53}. A simple, but accurate, numerical method
\cite{Hildebrand74} has been employed to evaluate the eigenvalues
in the domain $[0,1]$ in which the solutions are analytic.

Now returning to Eq.~(\ref{Time}) and taking $S(t)=e^{i\omega t}$,
we obtain the dispersion relation as
\begin{equation}
\omega_{ph}=\pm[\acute\lambda_{s}/2]^{1/2}\omega_{0}, \label{Sound}
\end{equation}
where $\acute\lambda_{s}$ is the smallest eigenvalues from Eq.
~(\ref{Last}).

The principal result of this paper is Eq.~(\ref{Sound}). It is
given in Fig.1, where the ratio $\omega_{ph}/\omega_{0}$ with
respect to the angular momentum $\ell$ is plotted. A close
examination of the dispersion curve shows that in the phonon
regime the energy spectrum of longitudinal collective excitations
is nearly linear with respect to $\ell$ (\textit{i.e.,} the wave
number $k_{\theta}\simeq \ell/r)$.

In an inhomogeneous medium, the Bogoliubov dispersion relation
$\omega_{ph}=ck$ with $c=[4\pi a \rho(r)\hbar^{2}]^{1/2}/M$ must
be studied approximately with $k_{\theta}\simeq\ell/r$ at a given
radius in SSNBEC. In particular $\omega_{ph}/\omega_{0}= 1.5972$
for $\ell=0$ is a unique value in a finite space problem, which
corresponds to a uniform radial perturbation. Finally, since the
eigenvalues are positive for all values of $\ell$, the Landau
critical velocity $v_{c}=\varepsilon_{ph}/p $ is also finite.
Hence the imperfect Bose gas is a superfluid, whereas an ideal
Bose gas, for which $c$ is independent of the scattering length,
is not, because $c$ and $v_{c}$ vanish identically. Moreover, the
dispersion relation, Fig.1, also shows how the presence of the
repulsive pair-interaction determines the excitation spectrum of
phonons, which is nearly linear with respect to $k_{\theta}$
\cite{Bog47, Fetter71}.

As we have shown above, the symmetry of the ground state wave
function breaks down at the free surface, {\it i.e.,} a nodal
surface. It is therefore natural to identify the underlying basic
mechanism for the symmetry breaking as \textit{a spontaneously
broken gauge symmetry} at the free surface which accompanies
phonons as Nambu-Goldstone bosons \cite{BenLee73,Weinberg96}.

Next we discuss a possible formation of a close-packed lattice as
$a\rightarrow 0$ in BEC. Our ground state wave function describes
a perfect Bose gas modified as little as possible by the presence
of the repulsive pair-interactions (a hard sphere approximation).
Hence the hard spheres in BEC could collapse to a close-packed
lattice in the limit, ({\it i.e.}, $a\rightarrow 0$) in which each
particle vibrates about a fixed center. Since the number of
particles $N$ in the system must be conserved in the limit, the
density must satisfy $(4/3)\pi d_{cl}^{3}\rho_{cl}=N$; $\rho_{cl}$
is the mean density of the collapsed spheres \cite{Dyson57,Lee57}.
It is unlikely that the closed-packed lattice will be realized in
a trap, because the radius of the condensate $b\rightarrow
d_{cl}\ll 1$ as $a\rightarrow 0$ [Eq.~(\ref{Eigen})], which
implies a far stronger external trapping force to overcome the
inter-particle repulsive force. However our analysis may provide a
qualitative mechanism by which the close-packed lattice can be
formed from BEC by symmetry breaking.

The separation of the zero-point vibration is not always possible,
but in the limit $a\rightarrow 0$, $b=\alpha=[8\pi
a\hbar^{2}\rho(0)]^{1/2}/(M\omega_{0})\rightarrow d_{cl}$,
$\acute{\lambda_{n}}\simeq 0$, we have the following zero-point
vibrational motion about fixed points. Since
$\lambda_{n}=5.0\omega_{0}^{2}$, we have a dispersion relation for
the closed-packed lattice from Eq.~(\ref{Time}),

\begin{equation}
\omega_{latt}=(5)^{1/2}\omega_{0}, \label{Lattice}
\end{equation}
which is again independent of $\hbar$. This implies that if there
is actually a transition from BEC to a crystalline ground state,
it is a classical lattice. It is gratifying to note that the
condition under which the close-packed lattice could be realized
in BEC is essentially the same as that of a crystalline lattice
observed under an external high pressure applied to He II
\cite{Mezhov66,Chan04}. The physical mechanism by which the
close-packed lattice can be realized is again the spontaneously
broken gauge symmetry which accompanies phonons as Nambu-Goldstone
bosons \cite{BenLee73,Weinberg96}.

In summary, we have shown the perturbation method with the
Lagrangian displacement vectors on particle trajectories in the
many-system yields the results that are consistent with those of
quantum field theory technique of Bogoliubov. The compressible
perturbations give the dispersion relation for a spherical
(longitudinal) sound wave which is of the form of the Bogoliubov
spectrum for phonons. The incompressible perturbations drive the
free surface to behave like a normal fluid. The dispersion
relations Eqs.~(\ref{Surf2}) and (\ref{Sound}) are interpreted as
the result of the spontaneously broken gauge symmetry at the free
surface which accompanies phonons (Nambu-Goldstone bosons). The
present approach gives a unified picture of the collective
excitations in BEC in a trap and also suggests a possible
mechanism by which BEC reduces to a close-packed lattice under
external pressure similar to that of the helium crystals at
constant pressure.

{\it Note added in proof}. After this paper was written, I learned
about 't Hooft's conjecture that is essentially equivalent to our
picture of fluctuation-dissipation of the longitudinal sound waves
at the free surface, which is driven by the effective quantum
potential and is due to the broken symmetry [Massimo Blasone, Petr
Jizba, Giuseppe Vitiello, Phys. Lett. A 287 (2001) 205].

\newpage
\begin{figure}
\includegraphics[scale=0.5]{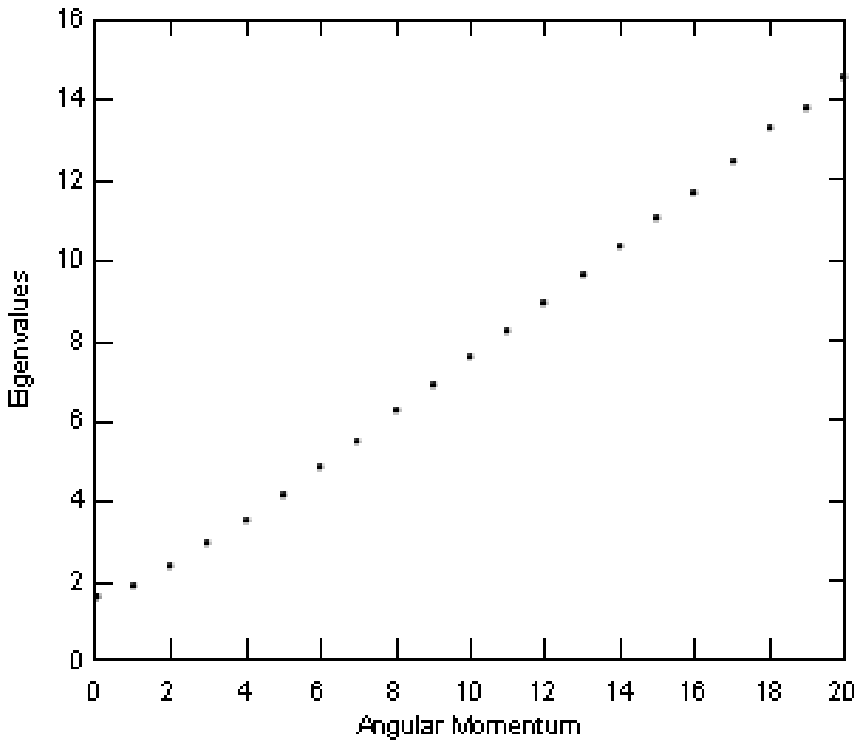}
\caption{The ratio $\omega_{ph}/\omega_{0}=
[2\acute\lambda_{s}]^{1/2}$ is plotted against the angular
momentum $\ell$. It shows how the energy spectrum of phonons
varies with the angular momentum.}%
\label{fig:fig1}
\end{figure}
%\end{widetext}

\begin{thebibliography}{95}
\bibitem{Lieb02} Elliott H. Lieb and Robert Seiringer, Phys. Rev.
        Lett. 88 (2002) 170409.%\marginpar{Lieb02}
\bibitem{Bog47} N. Bogoliubov, J. Phys. (USSR), 11 (1947) 23.
        %\marginpar{B47}
\bibitem{Penrose56} Oliver Penrose and Lars Onsager, Phys. Rev.
        104 (1956) 576.%\marginpar{Penrose56}
\bibitem{Feynman53} R. P. Feynman, Phys. Rev. 91 (1953) 1291, 1301; {\it ibid}
         94 (1954) 262. %\marginpar{Feynman53}
\bibitem{Landau41} L. D. Landau, J. Phys. (USSR), 5 (1941) 71; {\it
        ibid} 11 (1947) 91 (1947). %marginpar{Landau41}.
\bibitem{note1} The conventional perturbation methods are applicable
        to a periodic or a uniform system. In particular,
        the Bogoliubov theory is applicable only to a uniform
        medium, because one can quantize the scalar field of
        quasi-particles only in a Hilbert space; see James D. Bjorken and
        Sidney D. Drell, {\it Relativistic Quantum Fields} (McGraw Hill,
        1965) Chapt 11 and Gregor Wentzel, {\it Quantum Theory of Fields}
        (Dover Pulications, 1949) Chapt II.
\bibitem{Fetter01} Alexander L. Fetter and Anatoly A. Svidzinsky,
        J. Phys. Condens. Matter 13 (2001) R135. %\marginpar{Fetter01}
\bibitem{Lee57} T. D. Lee, K. Hwang, and C. N. Yang, Phys. Rev. {\bf106}, 1135
        (1957). %\marginpar{Lee57}
\bibitem{Steinhauer02} J. Steinhauer, {\it et al.}, Phys. Rev. Lett. 88 (2002)
        120407.%marginpar{Steinhauer02}
\bibitem{Anderson66} P. W. Anderson, Rev. Mod. Phys. 38 (1966) 249, and
        the references therein. %\marginpar{Anderson66}
\bibitem{Anderson62} P. W. Anderson, Phys. Rev. 130 (1962) 439.
        %\marginpar{Anderson62}
\bibitem{BenLee73} Ernest S. Abers and Benjamin W. Lee, Phys. Rep.
         9 (1973) 1 (1973).%\marginpar{BenLee73}
\bibitem{Weinberg96} Steven Weinberg, {\it The Quantum Theory of Fields}
        (Cambridge University Press, Cambridge 1996)
        Vol II Chapters 19 and 21.%\marginapar{Weinberg96}
\bibitem{Bohm52} David Bohm, Phys. Rev. 85 (1952) 166; 85 (1952)
        180.%\marginpar{Bohm52}
\bibitem{Aharonov63} Y. Aharonov and D. Bohm, Phys. Rev.
        130 (1963) 1625.%\marginpar{Aharonov}
\bibitem{Lieb98} Elliott H. Lieb and Jakob Yngvason, Phys. Rev.
        Lett. 80 (1998) 2504. %\marginpar{Lieb98}
\bibitem{BFKK58} I. Bernstein, E. Frieman, M. Kruskal, and R.
        Kulsrud, Proc. Royal Soc. (London) A  244 (1958) 17;
        Freeman J. Dyson,  J. Math. and Mech. 18 (1968) 91. %\marginpar{BFKK58}
\bibitem{Weinberg67} Steven Weinberg, J. Math. Phys. 8 (1967) 617.
        %\marginpar{67}
\bibitem{SJHan82} S. J. Han and B. R. Suydam, Phys. Rev. A  26 (1982)
        926.%\marginpar{SJHan82}
\bibitem{SJHan2} S. J. Han, cond-mat/0505373.%\marginpar{SJHan2}
\bibitem{Anderson58} P. W. Anderson, Phys. Rev. 112 (1958) 1900.
         %\marginpar{Anderson58}
\bibitem{Oliva89} J. Oliva, Phys. Rev. B
         39 (1989) 4197. %\marginpar{Oliva89}
\bibitem{note2} Because of the presence of $g\rho$ term in
        $V(\bm{x})$, Bohm's tentative interpretation of $\rho$ as
        a field coordinate and $S(\bm {x})$ as the momentum,
        canonically conjugate to $\rho$ is no longer valid.
\bibitem{SJHan1} S. J. Han, Phys. Rev. A 44 (1991) 5784.%\marginpar{SJHan1}
\bibitem{Landau59} L. D. Landau and E. M. Lifshitz, {\it Fluid
        Mechanics} (Pergamon Press, New York, 1959) Chapters I, VII, and VIII.
        %\marginpar{Landau59}
\bibitem{Lamb45} Sir H. Lamb, {\it Hydrodynamics} (Dover, New York,
        1945) Chapter IX. %\marginpar{Lamb}
\bibitem{Reif64} G. W. Rayfield and F. Reif, Phys. Rev. 136 (1964) A1194.
        %\marginpar{Reif64}
\bibitem{Glaberson68} W. L. Glaberson, D. M. Strayer, R. J. Donnelly, Phys. Rev.
        Lett. 20 (1968) 1428 ; {\it ibid} 21 (1968) 1740. %\marginpar{Glaberson68}
\bibitem{Osborne50} D. V. Osborne, Proc. Roy. Soc. (London) A 63 (1950) 909
\bibitem{Meservey64} R. Meservey, Phys. Rev. 133 (1964) A1471; ; see also
        P. L. Elliot, C. I. Pakes, L. Skbrek, and W. F. Vinen,
        Phys. Rev. B 61 (2000) 1396. %\marginpar{Meservey64}
\bibitem{Mott49} N. F. Mott, Phil. Mag. 40 (1949) 61. %\marginpar{Mott49}
\bibitem{Kubo57} R. Kubo, J. of Phys. Soc. Jpn. 12 (1957) 570;
         J. M. Luttinger, Phys. Rev.  135 (1964) A1505 .  %\Marginpar{Kubo57}
\bibitem{WW63} E. T. Whittaker and G. N. Watson, {\it A course
        of Modern Analysis}, Fourth Edition (Cambridge Press,
        1963) Chapter 14 and 23.%\marginpar{WW63}
\bibitem{Hildebrand74} F. B. Hildebrand, {\it Introduction to
         Numerical Analysis} (McGraw-Hill,1974),
         Second edition, Chapter 6.%\marginpar{Hildebrand}
\bibitem{Fetter71} A. L. Fetter and J. D. Wallecka, {\it Qunatum
        Theory of Many-Particle Systems} (MacGraw-Hill, 1971) Chapt. 14.
\bibitem{Dyson57} F. J. Dyson, Phys. Rev. 106 (1957) 20.
        %\marginpar{Dyson57}
\bibitem{Mezhov66} L. P. Mezhov-Deglin, Sov. Phys. JETP 22 (1966)
        47.%\marginpar{Mezhov66}
\bibitem{Chan04} E. Kim and M. H. W. Chan, Nature 427 (2004) 225;
        see also John M. Goodkind, Phys. Rev. Lett. 89 (2002)
        95301. %\marginpar{Chan04}


\end{thebibliography}
\end{document}